\documentclass[a4paper,11pt]{article}
\usepackage{jheppub,verbatim,mathrsfs}

\title{Hamilton-Jacobi Renormalization for Lifshitz Spacetime}

\author{Marco Baggio,}
\author{Jan de Boer}
\author{and Kristian Holsheimer}

\affiliation{Institute for Theoretical Physics, University of Amsterdam,\\
Science Park 904, Postbus 94485, 1090 GL Amsterdam, The Netherlands}

\emailAdd{m.baggio@uva.nl}
\emailAdd{j.deboer@uva.nl}
\emailAdd{k.holsheimer@uva.nl}

\abstract{Just like AdS spacetimes, Lifshitz spacetimes require counterterms in order
to make the on-shell value of the bulk action finite. We study these counterterms
using the Hamilton-Jacobi method. Rather than imposing boundary conditions from
the start, we will derive suitable boundary conditions by requiring that 
divergences can be canceled using only local counterterms. 
We will demonstrate in examples that this procedure indeed leads to a finite
bulk action while at the same time it determines the asymptotic behavior of
the fields. This puts more substance to the belief that Lifshitz spacetimes 
are dual to well-behaved field theories. As a byproduct, we will
find the analogue of the conformal anomaly for Lifshitz spacetimes.}

\keywords{Gauge-gravity correspondence, Holography and condensed matter physics (AdS/CMT), Anomalies in Field and String Theories}
\arxivnumber{1107.5562}

\begin{document}
\maketitle


\section{Introduction}

Lifshitz spacetimes were originally introduced as possible holographic dual
descriptions of non-relativistic field theories \cite{KLM,Koroteev:2007yp} and have since
appeared in many different setups, for example as IR geometries \cite{HT2}. Moreover,
they have appeared as solutions of string theory \cite{LST0,LST1,LST6,LST5,LST4,LST2,LST3} and although they are not yet at the same footing as ordinary AdS spacetimes, it is worthwhile to
explore to what extend the usual AdS/CFT techniques can be applied to 
Lifshitz spacetimes as well. 

Certain features of Lifshitz spacetimes that have been and still are 
confusing are its global causal structure, the absence of a version of
``global Lifshitz'', the nature of the boundary conditions on the metric
and other fields, and indications coming for example from Schr\"odinger holography \cite{Guica:2010sw} that one needs non-local counterterms
to remove divergences in the on-shell value of the action.

Motivated by this we decided to explore the nature of the divergences that
appear in Lifshitz spacetimes when computing the on-shell value of the effective
action using the Hamilton-Jacobi method, which turns out to be more efficient
in this case than using the Fefferman-Graham expansion, which rapidly becomes
quite intractable. 

Normally, in order to perform holographic renormalization, one needs to
first say something about the boundary conditions for the fields. We will, however,
follow a different and novel approach. As we will show, if we require that all divergences
should be canceled by local counterterms, this will automatically enforce
particular boundary conditions for the fields. More precisely, we will find
that particular local covariant quantities made out of the bulk fields
have to scale in a specific way as we approach the boundary of Lifshitz.
With this approach, we will also show that for a class of bulk Lagrangians all 
power law divergences
can indeed be canceled using only local counterterms. This strongly suggests that Lifshitz
spacetimes are dual to field theories with a well-defined UV completion.

In addition, certain ambiguities that appear in the analysis of the counterterms 
have a natural interpretation in the dual field theory in terms of marginal 
deformations, exactly as was the case for AdS/CFT.

Along the way, we will show that counterterms that had been previously
proposed in \cite{RS} are insufficient to cancel divergences beyond the 
leading order, and find the analogue of the conformal anomaly for Lifshitz
spacetimes. 

The outline of this paper is as follows. In section 2 we review the Hamilton-Jacobi
method and apply it to the non-derivative terms in the boundary effective
action. All power-law divergent terms in the effective action can be canceled
using local counterterms. Sometimes, logarithmically divergent terms appear which cannot
be canceled using local counterterms, and it is precisely these that are responsible 
for the analogue of the conformal anomaly. We also describe the relation between
ambiguities that appear and the existence of marginal deformations. 

In section 3 we perform a non-trivial consistency check by explicitly
computing the on-shell action for scalar perturbations of the metric and
gauge field to second order. We will find that with our counterterms the
on-shell action is indeed rendered finite. 

Various subtleties, such as the presence of logarithmic divergences, qualitative
dependence of the answers on the value of the so-called dynamical exponent $z$, and issues related to the boundary
conditions are discussed in the conclusions.

The appendices contain some background material and a brief description of
the extension of our methods to the terms containing derivatives.

{\bf Note added:} As we were preparing this paper for submission to the ArXiv, 
the paper \cite{Ross2011} appeared, which reaches similar conclusions as we do
though using different methods. 


\section{Holographic Renormalization}\label{sec:hj}
In this section we set up the general framework for computing the counterterm action. We begin with a brief review of Lifshitz spacetime and the specific bulk action we shall use. After that, we describe the Hamilton--Jacobi method of holographic renormalization and we introduce the `Lifshitz scaling anomaly'. Finally, we explicitly compute the counterterms at the level of no spacetime derivatives and contributions to the Lifshitz scaling anomaly. We carry out the analysis adding a scalar field, which makes discussions of several issues particularly transparent.

\subsection{Lifshitz spacetime and the Einstein--Proca action}
\label{sec:lsatepa}
Lifshitz spacetime is a proposed gravitational dual to a field theory at a UV fixed point with anisotropic (Lifshitz-like) scaling symmetry,
\begin{align}\label{eq:lif-scaling}
\begin{pmatrix}x\\t\end{pmatrix}\mapsto\begin{pmatrix}\lambda\,x\\\lambda^zt\end{pmatrix}.
\end{align}
The configuration of $(d+1)$-dimensional Lifshitz spacetime \cite{KLM} that we consider consists of the following metric and vector \cite{Taylor},
\begin{equation}\label{eq:lifshitz-solution}
ds^2\ =\ dr^2-e^{2zr}dt^2+e^{2r} d\vec x{\,}^2,\qquad\qquad A=\sqrt{-\alpha_0}\,e^{zr} dt.
\end{equation}
This metric is invariant under the so-called Lifshitz algebra \cite{Adams:2008zk}, which consists of time translations, spatial translations, spatial rotations, and anisotropic scaling invariance \eqref{eq:lif-scaling} (with a simultaneous shift in the radial coordinate $r\mapsto r-\log\lambda$). Unlike so-called Schr\"odinger spacetimes, the Lifshitz spacetime is \emph{not} invariant under Galilean boosts $x\mapsto x+vt$. We will eventually work in 3+1 bulk spacetime dimensions ($d=3$), but we keep $d$ arbitrary for as long as possible. These fields comprise a solution to the Einstein--Proca action $S=S_\text{grav}+S_{A}$, with
\begin{align}
\label{eqn:action1}
S_\text{grav} &=\int d^{d+1}x\sqrt{-g}\left(R-2\Lambda\right)+\int d^{d}\xi\sqrt{-\gamma}\ 2K,\\
\label{eqn:action2}
S_A		&=\int d^{d+1}x\sqrt{-g}\left(-\frac14F_{\mu\nu}F^{\mu\nu}-\frac12m^2A_\mu A^\mu\right),
\end{align}
where we used the convention $16\pi G = 1$. It should be noted that we could also have chosen a different action that has the Lifshitz metric as a solution, see e.g.\ \cite{KLM,Taylor}. We have chosen the Einstein--Maxwell theory for its relative simplicity. In order to find Lifshitz spacetime as a solution, we must pick our parameters to be
\begin{equation}\label{eq:lifshitz-parameters}
\Lambda=-\frac12(z^2+z+4), \qquad\qquad m^2=2z, \qquad\qquad \alpha_0 = -2\frac{z-1}{z}.
\end{equation}
The equations of motion are
\begin{align}
 R_{\mu\nu} -\frac{1}{2}R g_{\mu\nu} + \Lambda g_{\mu\nu}  &= \frac12T_{\mu\nu},   &
 \nabla_{\mu} F^{\mu\nu} &= m^{2} A^{\nu},
\end{align}
where $T^{\mu\nu}=\frac2{\sqrt{-g}}\frac{\delta S_A}{\delta g_{\mu\nu}}$ is the Proca stress tensor. We wish to add a scalar at some point, so let us give the scalar action as well,
\begin{align}
\label{eqn:scalaraction}
S_\phi = \int d^{d+1}x\sqrt{-g} \left( -\frac12\partial_\mu\phi\partial^\mu\phi-V(\phi) \right),
\end{align}
with a potential $V(\phi)=\frac12\mu^2\,\phi^2+v_3\phi^3+v_4\phi^4+...$, which is presumed to be known. It is in principle possible to consider different setups, e.g.\ with direct couplings between the scalar field and the vector field such as $\phi^{2} A^{2}$, but for simplicity we will restrict our attention to the simple case \eqref{eqn:scalaraction}.

At this point it should be stressed that adding a scalar is of secondary importance when it comes to the main goal of this work, which is to show that the on-shell action can be renormalized by adding local counterterms alone.\footnote{This is true for $1<z<2$; when $z>2$ there will be a mode that acts as a source for an irrelevant operator in the field theory and could in principle give a divergence that must be removed by adding  non-local counterterms along the lines of \cite{vanRees:2011fr}. This will be discussed in more detail in due time below.} The reason for including the scalar $\phi$ nonetheless is twofold. First, we shall find that it is convenient to consider the composite scalar $\alpha=\gamma^{ab}A_aA_b$. In order to have a nice intuitive understanding of the radial scaling behavior of the quantity $\alpha$, we include the scalar field $\phi$ for relative comparison. In particular, section \ref{sec:radial-behavior} is devoted to the dicussion of the radial scaling of $\alpha$ (and $\phi$) and section \ref{sec:origin-ambiguities} explains possible ambiguities in solving the Hamilton--Jacobi equation and their relation to an anisotropic version of the holographic Weyl anomaly. The second reason for including $\phi$ in our discussion is that we can illustrate explicitly that we may find anomalous breaking of the symmetry under anisotropic scaling transformations in a simple setting. We also expect such anomalous symmetry breaking in the pure Lifshitz background, i.e.\ without the scalar field, which will be explored in future work.

Our aim is to construct a finite on-shell action for an appropriate class of asymptotically Lifshitz spacetimes, where this notion will be made more precise in the following. As usual, the action given in \eqref{eqn:action1} and \eqref{eqn:action2} diverges on-shell, and it is necessary to introduce a set of counterterms to remove these divergences. In this paper we will assume that such counterterms are local in the fields, and we will use the Hamilton-Jacobi method to determine their form. We expect the on-shell action to be of the form
\begin{equation}
\label{eqn:splitlocgamma}
S_\text{cl}=S_{\text{loc}} + \Gamma,
\end{equation}
where $S_{\text{loc}}$ contains the local power-law divergent terms and $\Gamma$ diverges at most logarithmically. We will determine $S_{\text{loc}}$ by imposing the Hamilton constraint $H=0$, where $H$ is the radial Hamiltonian corresponding to the Einstein--Proca action, which is derived in appendix \ref{sec:hamiltonian}; it is given by
\begin{align}
H = \int_{\Sigma_r}d^dx\sqrt{-\gamma} \left( N\mathcal{H}+N^a\mathcal{H}_a \right),
\end{align}
where $\Sigma_r$ is a hypersurface of large but constant $r$, while $N$ and $N^a$ are the usual lapse and shift functions. The momentum constraint is given by
\begin{align}
\mathcal{H}_a = -2D^b\pi_{ab}-A_aD_bE^b+F_{ab}E^b+\pi \partial_a\phi = 0,
\end{align}
where the quantities $\pi^{ab}$, $E^a$ and $\pi_\phi$ are the canonical momenta dual to the induced metric $\gamma_{ab}$, induced vector $A_a$ and the scalar $\phi$ respectively. The Hamiltonian constraint is
\begin{align}
\label{eqn:hamconstr}
\mathcal{H}=-\left(\pi_{ab}\pi^{ab}-\frac1{d-1}\pi^2\right)-\frac12E^aE_a-\frac12\pi_\phi^2-\frac1{2m^2}(D_aE^a)^2-\mathcal{L} = 0,
\end{align}
where $\mathcal{L}=R-2\Lambda-\frac14F_{ab}F^{ab}-\frac12m^2A_aA^a-\frac12\partial_a\phi\partial^a\phi-V(\phi)$ is the Lagrangian restricted to $\Sigma_r$.


\subsection{Hamilton--Jacobi equation and the Lifshitz-scaling anomaly}
\label{sec:hja}
The Hamilton--Jacobi (HJ) equations of motion for a point particle are $H=-\partial_t S_\text{cl}$ and $p(t)=\partial S_\text{cl}/\partial q$, where the on-shell action $S_\text{cl}$ is the action evaluated on the classical path with given initial and final conditions. The first HJ equation simply becomes $H=0$, while the second one  is generalized to
\begin{align}\label{eq:2nd-hj-eq}
\pi^{ab}(r)\ &=\ \frac1{\sqrt{-\gamma}}\frac{\delta S_\text{cl}}{\delta \gamma_{ab}}(r), &
E^a(r)\ &=\ \frac1{\sqrt{-\gamma}}\frac{\delta S_\text{cl}}{\delta A_a}(r), &
\pi_\phi(r)&=\frac1{\sqrt{-\gamma}}\frac{\delta S_\text{cl}}{\delta \phi}(r).
\end{align}
The two HJ equations may be combined into what is known as \emph{the} Hamilton--Jacobi equation,
\begin{align}
H\big(\gamma_{ab},A_a,\phi;\tfrac{\delta S_\text{cl}}{\delta \gamma_{ab}},\tfrac{\delta S_\text{cl}}{\delta A_a},\tfrac{\delta S_\text{cl}}{\delta \phi}\big)=0.
\end{align}
The HJ equation is a functional PDE for the on-shell action. In principle, this equation determines the form of the on-shell action $S_{\text{cl}}$, but it is far too difficult to solve. Since we are interested only in the local part $S_{\text{loc}}$, we can recast the problem in a more tractable form.
It is useful to introduce the following notation for the `kinetic' part of the Hamiltonian constraint \eqref{eqn:hamconstr}:
\begin{align}\label{eq:hj-brackets}
(\sqrt{-\gamma})^2\,\{F,G\}&\equiv-\left(\gamma_{ac}\gamma_{bd}-\frac1{d-1}\gamma_{ab}\gamma_{cd}\right)\frac{\delta F}{\delta \gamma_{ab}}\frac{\delta G}{\delta \gamma_{cd}}\\
&\hspace{3cm}-\frac12\frac{\delta F}{\delta \phi}\frac{\delta G}{\delta \phi}-\frac12\gamma_{ab}\frac{\delta F}{\delta A_a}\frac{\delta G}{\delta A_b}-\frac1{2m^2}D_a\frac{\delta F}{\delta A_a}\,D_b\frac{\delta G}{\delta A_b}.\nonumber
\end{align}
such that the Hamiltonian constraint is simply
\begin{align}
0=\mathcal{H}=\{S_\text{cl},S_\text{cl}\}-\mathcal{L}.
\end{align}
The bracket $\{F,G\}$ is symmetric and bilinear in $F$ and $G$. Therefore we can use the splitting $S_\text{cl}=S_{\text{loc}} + \Gamma$ to write:
\begin{equation}\label{eq:HJ-full}
0 = \{S_\text{loc},S_\text{loc}\}-\mathcal{L} + 2 \{S_\text{loc},\Gamma \} + \{\Gamma,\Gamma\}.
\end{equation}
We define the ``local part'' of this expression as $\mathcal{H}_\text{loc} \equiv \{S_\text{loc},S_\text{loc}\}-\mathcal{L}$. The divergent part of $\mathcal{H}_\text{loc}$ should vanish by itself because the non-local part shouldn't contain power-law divergences. Solving $\mathcal{H}_\text{loc,div}=0$ determines the divergent terms in $S_\text{loc}$, which will be our counterterms.

We will see that, in the presence of marginal deformations, this procedure possibly leaves a finite remainder in $
\mathcal{H}_\text{loc}$, which we denote by $\mathcal{H}_\text{rem}$ and is generically determined unambiguously. Let us now rewrite the other piece in \eqref{eq:HJ-full} (for simplicity we assume $\gamma_{ti}=0$),
\begin{align}
2\int d^dx\sqrt{-\gamma} \{S_\text{loc},\Gamma \}\ &=\ \int d^dx\, \left( \dot\gamma_{ab}\frac{\delta \Gamma}{\delta \gamma_{ab}}+\dot A_a\frac{\delta \Gamma}{\delta A_a}+\dot\phi\frac{\delta \Gamma}{\delta \phi} \right) \\
&=\ \int d^dx\, \left( 2z\hat\gamma_{tt}\frac{\delta \Gamma}{\delta \hat\gamma_{tt}}+2\hat\gamma_{ij}\frac{\delta \Gamma}{\delta \hat\gamma_{ij}}+z\hat A_t\frac{\delta \Gamma}{\delta\hat A_t} + \lambda^{-}_{\phi} \, \hat\phi \frac{\delta \Gamma}{\delta \hat\phi} \right)+\ldots \\
&=\ \int d^dx\sqrt{-\hat\gamma}\,\Big( z\langle T_t{}^t\rangle+\langle T_i{}^i\rangle+z\hat A_t\langle \mathcal{O}_{\!A}\rangle  + \lambda^{-}_{\phi}\, \hat\phi \langle \mathcal{O}_{\!\phi}\rangle\Big)+\ldots,
\end{align}
which holds in the large-$r$ limit and where we used the hatted notation for the asymptotic values of the fields, e.g.\ $\hat\gamma_{tt}=\lim_{r\rightarrow\infty} e^{-2zr}\gamma_{tt}$, $\lambda^{-}_{\phi}$ is the leading radial scaling of the scalar field and the dots represent subleading contributions. We also used
\begin{align}
\langle T^{ab}\rangle&=\frac2{\sqrt{-\hat\gamma}}\,\frac{\delta \Gamma}{\delta \hat\gamma_{ab}}, &
\langle\mathcal{O}_{\!A}\rangle &=\frac1{\sqrt{-\hat\gamma}}\,\frac{\delta \Gamma}{\delta \hat A_t}, &
\langle\mathcal{O}_{\!\phi}\rangle &=\frac1{\sqrt{-\hat\gamma}}\,\frac{\delta \Gamma}{\delta \hat \phi}.
\end{align}
The final piece in \eqref{eq:HJ-full}, $\{\Gamma,\Gamma\}$, is subleading and vanishes in the large-$r$ limit. Plugging everything back into \eqref{eq:HJ-full} we immediately arrive at the Lifshitz analogue of the Weyl anomaly \cite{HS,BVV,KMM,MM}:
\begin{align}\label{eq:lif-anomaly}
\mathcal{A}_z\ \equiv\ z\langle T_t{}^t\rangle+\langle T_i{}^i\rangle+z\hat A_t\langle \mathcal{O}_{\!A}\rangle\ + \lambda^{-}_{\phi}\, \hat\phi \langle \mathcal{O}_{\!\phi}\rangle =\ \lim_{r\rightarrow\infty}e^{(z+2)r}\mathcal{H}_\text{rem}.
\end{align}
 Recall that $\mathcal{H}_\text{rem}$ was defined to be the finite remainder in $\mathcal{H}_\text{loc}$. We thus see that there is a possibility that the Lifshitz scaling symmetry is broken. We shall find that this happens for some specific values of the scalar mass-squared $\mu^2$. It should be noted that the quantity $T^a{}_b$ is \emph{not} the stress tensor of the (proposed) dual field theory; for a discussion of this subtlety, see e.g.\ \cite{Mann:2011hg} in which the tensor $T^a{}_b$ is compared to the so-called stress tensor \emph{complex} defined in \cite{RS,Ross2011}.


\subsection{Initial conditions}\label{acbycs}

As we mentioned earlier, one typically chooses an Ansatz that is covariant  \cite{BVV,MM,KMM}, such that the momentum constraint is automatically satisfied. Solving the HJ equation thus reduces to solving the Hamiltonian constraint. In the present case, the most general covariant Ansatz one can take is
\begin{align}
S_{\text{loc}}\ =\ \int_{\Sigma_r}d^dx\sqrt{-\gamma}\,U(\alpha,\phi)+\text{(derivative terms)}.
\end{align}
The quantity $\alpha\equiv A_a A^a$ is the only scalar that one can construct from the metric and the vector field containing no derivatives. At present, our main focus will be on this level of no spacetime derivatives; see Appendix \ref{sec:higher-derivatives} for a discussion of the higher-derivative levels.

As an aside, notice that the induced metric may be viewed as the metric on $d$-dimensional flat space with an $r$-dependent speed of light $c_r=e^{(z-1)r}$  \cite{RS}. More specifically,
\begin{equation}
\gamma_{ab}(r)\, dx^a dx^b\ =\ e^{2r}\left( -c^2_r\,dt^2+d\vec x{\,}^2 \right).
\end{equation}
Therefore, heuristically speaking, it seems quite natural to impose covariance on $\Sigma_r$; it is only in the limit $r\rightarrow\infty$ that the speed of light becomes infinite.

Before we move on to solving the HJ equation, we need to establish the leading-order behavior of $U(\alpha,\phi)$ in an expansion about the Lifshitz background. In fact the HJ equations are functional differential equations for $S_{\text{cl}}$, and in order to solve them it is necessary to provide the initial conditions, that is the value of $S_{\text{cl}}$ for a solution of our choice. Henceforth, we set $d=3$. Recall that the background value of $\alpha$ is given by $\alpha_0=-2(z-1)/z$, cf.\eqref{eq:lifshitz-parameters}. We shall use the Hamilton equation of the type $\dot{q}=\partial H/\partial p$ in order to fix the leading-order behavior of $U(\alpha,\phi)$.
\begin{align}
\partial_r\gamma_{ab}\ &=\ \frac{\delta H}{\delta(\sqrt{-\gamma}\,\pi^{ab})}\ =\ -2\pi_{ab}+\gamma_{ab}\pi, \\
\partial_r A_a\ &=\ \frac{\delta H}{\delta(\sqrt{-\gamma}\,E^a)}\ =\ -E_a+\frac1{m^2}D_aD_bE^b,\\
\partial_r \phi\ &=\ \frac{\delta H}{\delta(\sqrt{-\gamma}\,\pi_\phi)}\ =\ -\pi_\phi.
\end{align}
At the level of no spacetime derivatives, the canonical momenta are given (via the second HJ equation \eqref{eq:2nd-hj-eq}) by:
\begin{align}
\pi_{ab}\ &=\ \frac12\gamma_{ab}U-A_aA_b\,\frac{\partial U}{\partial \alpha}, \\
E_a\ &=\ 2A_a\,\frac{\partial U}{\partial \alpha}, \\
\pi_\phi &= \frac{\partial U}{\partial \phi},
\end{align}
where now $\pi^{ab}=\frac1{\sqrt{-\gamma}}\frac{\delta S_\text{loc}}{\delta \gamma_{ab}}$ denotes only the local part of the momentum (and similarly for $E^{a}$ and $\pi_\phi$), so that we find
\begin{align}
\partial_r\gamma_{ab}\ &=\ \left(\frac12U-\alpha\,\frac{\partial U}{\partial \alpha}\right)\gamma_{ab}+2A_aA_b\,\frac{\partial U}{\partial \alpha} \label{eq:ham-eq-gamma},\\
\partial_r A_a\ &=\ -2A_a\,\frac{\partial U}{\partial \alpha} \label{eq:ham-eq-A},\\
\partial_r \phi\ &=\ -\frac{\partial U}{\partial \phi} \label{eq:ham-eq-phi}.
\end{align}
Let us expand the function $U(\alpha,\phi)$ around the Lifshitz background as
\begin{align}
U(\alpha,\phi)\ =\ \sum_{m,n}u_{mn}\,(\alpha-\alpha_0)^m\phi^n.
\end{align}
We can then translate the radial behavior of the Lifshitz background $\partial_r\gamma_{tt}=2z\gamma_{tt}$, $\partial_r\gamma_{ij}=2\gamma_{ij}$, $\partial_rA_t=zA_t$, and $\partial_r\phi=0$ ($\phi=0$ in the background) into the values for the first few coefficients in the function $U(\alpha,\phi)$. We evaluate the above Hamilton equation on the Lifshitz background, so that we find
\begin{align}\label{eq:lifshitz-initial-cond}
u_{00}&=2(z+1), & u_{10}&=-\frac{z}2, & u_{01}=0.
\end{align}
Therefore, the first three coefficients in the expansion for $U(\alpha,\phi)$ around the background have been determined by imposing that the background is a solution to the equations of motion.


\subsection{Radial behavior of $(\alpha-\alpha_0)$ and $\phi$}\label{sec:radial-behavior}
The radial behavior of $(\alpha-\alpha_0)$ and $\phi$ is dictated by the value of the coefficients $u_{20}$ and $u_{02}$ respectively. We will now show how this comes about.

Let us start with the more familiar example of the scalar field $\phi$. The radial behavior can be obtained directly from the free scalar field equation in the Lifshitz background,
\begin{align}
\label{eqn:eqnmotionscalar}
\partial_r^2\phi+(z+2)\partial_r\phi+\gamma^{ab}\,\partial_a\partial_b\phi=\mu^2\phi.
\end{align}
Taking the Ansatz $\phi=\chi(t,x,y)\exp(\lambda_\phi\,r)$ yields
\begin{align}
\lambda_\phi\ =\ -\frac12\left( (z+2)\pm\sqrt{(z+2)^2+4\mu^2} \right).
\end{align}
Alternatively, one can find this radial behavior by expanding the Hamilton equation \eqref{eq:ham-eq-phi} to second order, such that
\begin{align}
\label{eqn:drphi}
\partial_r\phi\ =\ -2u_{02}\,\phi+...
\end{align}
Later, we will find that $u_{02}=-\lambda_\phi/2$, which agrees with the answer provided by the equations of motion \eqref{eqn:eqnmotionscalar}. Notice that there is also the Lifshitz analogue of the Breitenlohner--Freedman bound, i.e.\ $\mu^2\geq-(z+2)^2/4$, see also \cite{KLM,Taylor}. 

For $(\alpha-\alpha_0)$, we similarly find via \eqref{eq:ham-eq-gamma} and \eqref{eq:ham-eq-A}
\begin{align}
\label{eqn:dralpha}
\partial_r(\alpha-\alpha_0)\ =\ -\frac12\alpha\, U-(\alpha^2+4\alpha)\frac{\partial U}{\partial \alpha}\ =\ \lambda_\alpha(\alpha-\alpha_0)+...,
\end{align}
with
\begin{align}
\label{eqn:radalpha}
\lambda_\alpha\ =\ \frac{8(z^{2}-1)}{z^{2}}u_{20} -\frac32(z-1).
\end{align}
In this case finding the radial behavior using the Einstein--Proca field equations would be considerably more difficult, since $\alpha$ is a composite field containing both the vector and the metric.

One may wonder whether the bilinear term in $U(\alpha,\phi)$ spoils these relations. Later on, we will find that the HJ equation gives $u_{11}=0$, so this term is in fact absent.


\subsection{Non-derivative counterterms}\label{sec:counterterms}
We shall now set out to find the counterterms at the level of no spacetime derivatives by solving the local part of the HJ equation. As we mentioned before, this comes down to solving the local part of the Hamiltonian constraint,
\begin{align}
0 & \cong \ \{S_\text{loc},S_\text{loc}\}-\mathcal{L} \\
&=\ \frac{3}{8} U^{2} - \Big(\frac{1}{2}\alpha^{2} + 2 \alpha\Big) \left(\frac{\partial U}{\partial \alpha}\right)^2  - \frac{1}{2}\alpha U \frac{\partial U}{\partial \alpha}-\frac12\left(\frac{\partial U}{\partial \phi}\right)^2 -(z^2+z+4) + z\alpha + V(\phi),
\end{align}
where the symbol $\cong$ indicates that there might be a finite remainder $\mathcal{H}_\text{rem}$ as discussed in section \ref{sec:hja}. Remember that $S_\text{loc}=\int d^dx\sqrt{-\gamma}\,U+\text{(derivatives)}$ and we expanded the function $U$ about the Lifshitz background as $U(\alpha,\phi)=\sum_{m,n}u_{mn}\,(\alpha-\alpha_0)^m\phi^n$. We can do a similar expansion of the non-derivative part of the local Hamiltonian constraint,
\begin{align}
\mathcal{H}_\text{loc,non-deriv}\ =\ \sum_{m,n}\mathcal{H}_{mn}\,(\alpha-\alpha_0)^m\phi^n.
\end{align}
This allows us to solve the Hamiltonian constraint order by order.

\textbf{Order zero and order one.}
The above values for $u_{00}$, $u_{10}$, and $u_{01}$ are such that both the order-zero and the order-one constraints vanish identically, i.e.\ $\mathcal{H}_{00}=\mathcal{H}_{01}=\mathcal{H}_{10}=0$, even though they depend on $u_{20}$ in principle.

\textbf{Order two.}
The constraints at second order in $(\alpha-\alpha_0)$ and/or $\phi$ are
\begin{align}
\mathcal{H}_{20}\ &=\  - \frac12 u_{11}^2 - (2z-5)u_{20} + 
 8\frac{z^2-1}{z^2} u_{20}^2 -\frac{5 z^2}{32},\\
\mathcal{H}_{11}\ &=\ \frac{1}{2}\left(8\frac{z^2-1}{z^2} u_{20}-2 u_{02}-z+7\right) u_{11}, \\
\mathcal{H}_{02}\ &=\ 2\frac{z^2-1}{z^2} u_{11}^2  + (z+2)u_{02} - 2 u_{02}^2 + \frac12\mu^2,
\end{align}
which is a system of three coupled equations in terms of the three unknowns $u_{20}$, $u_{11}$, and $u_{02}$. There is also one free parameter (apart from $z$), namely the scalar mass-squared $\mu^2$. The generic case, for which $\mu^2\neq2 (z-1)(z-2) $, one finds
\begin{align}
u_{20}&=-\frac{z^2}{16(z^2-1)}\left(5-2z\pm\sqrt{9z^2-20z+20}\right), \\
u_{11}&=0, \\
u_{02}&=\frac14 \left( z+2 \pm \sqrt{(z+2)^2 + 4 \mu^2}\right).
\end{align}
For $\mu^2=2(z-1)(z-2)$, however, one of the three coefficients $(u_{20},u_{11},u_{02})$ remains undetermined. In principle, it is possible that there is a remainder in the Hamiltonian constraint that cannot be set to zero by tuning a coefficient. Such a remainder would contribute to a Lifshitz scaling anomaly. Although, in the present case, for this specific value of $\mu^{2}$ we find no such remainder.

Remember from Section \ref{sec:radial-behavior} that the coefficients $u_{20}$ and $u_{02}$ determine the radial behavior for the fields $(\alpha-\alpha_0)$ and $\phi$ respectively. The choice of the sign in $u_{20}$ and $u_{02}$ correspond to choosing the either the normalizable or non-normalizable mode for the fields. For both $u_{20}$ and $u_{02}$, the minus-sign root will correspond to the non-normalized modes. Though, for the scalar field $\phi$ there is a special window for $\mu^2$ where both modes are normalizable and one has the freedom to choose the plus sign in $u_{02}$. This special window is given by $-\frac14 (z+2)^{2} \leq \mu^{2} < -\frac14 (2-z)(2+3z)$, which is obtained by requiring that the Klein--Gordon norm is finite.

\textbf{Order three.}
The constraints at third order are given by
\begin{align}
\mathcal{H}_{30}\ &=\ -\frac{1}{2} \big(z+2-3 \beta_z\big)u_{30} ,\\\nonumber
&\quad +\frac{z^3}{128(z^2-1)^2} \big(35 z^3-160 z^2+223 z-170+\left(15 z^2-30 z+39\right) \beta_z \big)\\
\mathcal{H}_{21}\ &=\ -\frac{1}{2} \big(z+2-\gamma_{z}-2 \beta_z\big)u_{21},\\
\mathcal{H}_{12}\ &=\ -\frac{1}{2}\big(z+2 - \beta_z-2\gamma_{z}\big)u_{12}
 +\frac{z\big( z-6+\beta_z \big) \big(z+2-\gamma_{z}\big)}{32(z+1)},\\
\mathcal{H}_{03}\ &=\ -\frac12\big(z+2 -3\gamma_z\big)u_{03}+v_3,
\end{align}
where we introduced the abbreviations for the square-roots $\beta_z=\sqrt{9z^2-20z+20}$ and $\gamma_{z}(\mu)=\sqrt{(z+2)^2+4\mu^2}$.\footnote{The quantity $\beta_z$ is positive for any value of $z$. Since $\gamma_z$ depends on $\mu^2$, we see that $\gamma_z>0$ as long as $\mu^2>-\frac14(z+2)^2$.} Recall that $v_{3}$ is the third order coefficient in the potential $V(\phi)$. The first thing one sees is that the equations decouple already at third order; this continues to hold at higher order (for as long as one does not hit a continuous ambiguity). One determines the coefficients $u_{30}$, $u_{21}$, $u_{12}$, and $u_{03}$ from these equations for generic values of $\mu^2$. There are two special values for which one of the coefficients remains undetermined.
\begin{itemize}
\item \textbf{Ambiguity 1.} For $\mu^2=-\frac29(z+2)^2$, which is above the BF bound, the coefficient $u_{03}$ remains undetermined. There is a remainder in the Hamiltonian constraint that cannot be set to zero, $\mathcal{H}_\text{rem}=v_3\,\phi^3$, there is a contribution to the Lifshitz-scaling anomaly \eqref{eq:lif-anomaly},
\begin{align}
\mathcal{A}_z\ =\ v_3\,\lim_{r\rightarrow\infty}\left(e^{(z+2)r}\,\phi^3\right)+\ldots
\end{align}
which is present for all values of $z$. This is the same contribution to the anomaly that was found in AdS ($z=1$), see e.g.\ \cite{Boer,MM}
\item \textbf{Ambiguity 2.} We find a second ambiguity when $\mu^2= \frac1{16}(z+2-\beta_z)^2-\frac14(z+2)^2$, the coefficient $u_{12}$ is undetermined and there is also a remainder, namely
\begin{align}
\mathcal{H}_{12}\ &=\ -\frac{z\big( z-6+\beta_z \big) \big(z+2+\beta_{z}\big)}{64(z+1)}.
\end{align}
Thus, for this specific value of $\mu^2$, so we have a contribution to the Lifshitz anomaly,
\begin{align}
\mathcal{A}_z\ =\ - \frac{z\big( z-6+\beta_z \big) \big(z+2+\beta_{z}\big)}{64(z+1)}\,\lim_{r \rightarrow\infty}\left(e^{(z+2)r}(\alpha-\alpha_0)\phi^2\right)+\ldots
\end{align}
Interestingly, this contribution to the Lifshitz anomaly vanishes for $z=2$.
\end{itemize}


\subsection{Origin of the continuous ambiguities}\label{sec:origin-ambiguities}
The continuous ambiguities arise when the radial behavior of a term mixing $(\alpha-\alpha_0)$ with $\phi$ becomes of $O(1)$. In order to see this, let us recall the radial behavior (at large $r$) from Section \ref{sec:radial-behavior}, i.e.\ $(\alpha-\alpha_0)\sim\exp(\lambda_\alpha\,r)$ and $\phi\sim\exp(\lambda_\phi\,r)$, where
\begin{align}
\lambda_\alpha\ &=\ \frac{8(z^{2}-1)}{z^{2}}u_{20} -\frac32(z-1), &
\lambda_\phi\ &=\ -2u_{02}.
\end{align}
When we plug in the values we found from the HJ equation, we get
\begin{align}
\lambda_\alpha^\pm\ &=\ -\frac12\left( z+2\,\pm\,\sqrt{(z+2)^2+8(z-1)(z-2)} \right), \\
\lambda_\phi^\pm\ &=\ -\frac12\left( z+2\,\pm\,\sqrt{(z+2)^2+4\mu^2} \right).
\end{align}
Let us refer to the corresponding modes as $(\alpha-\alpha_0)_\pm$ and $\phi_\pm$. For generic values of the scalar mass $\mu$, we see that $\lambda_{\alpha,\phi}^-$ describes the radial behavior of the non-normalizable modes (sources) and  $\lambda_{\alpha,\phi}^+$ that of the normalizable modes (vevs). In terms of the scaling coefficients, we always have
\begin{align}
\lambda_\alpha^-+\lambda_\alpha^+&=-(z+2), & \lambda_\phi^-+\lambda_\phi^+&=-(z+2).
\end{align}
Note that this $-(z+2)$ cancels against the $+(z+2)$ coming from $\sqrt{-\gamma}\sim e^{(z+2)r}$. 

\textbf{Order-two ambiguity.}
The critical value $\mu^2=2(z-1)(z-2)$ that we found at second order in $(\alpha-\alpha_0)$ and $\phi$ is simply a result of tuning the scalar mass such that $\lambda_\phi^- =\lambda_\alpha^-$. We should stress that this ambiguity is not related to the presence of a marginal operator, therefore it is not surprising that there is no remainder in this case. This ambiguity is parametrized by $u_{11}$, which mixes $\alpha - \alpha_{0}$ and $\phi$ by giving linear contributions to the right hand side of \eqref{eqn:drphi} and \eqref{eqn:dralpha}. This suggests the possible presence of a one-parameter family of allowed boundary conditions, but we leave this for further study.

\textbf{Order-three ambiguities.}
At third order, we found two ambiguities, which comes from the freedom to tune the scalar mass such that either the combination $\phi^3$ or $(\alpha-\alpha_0)\phi^2$ is of order one in $e^{-(z+2)r}$. Another way of saying this is that either the coefficient $u_{03}$ or $u_{12}$ remains undetermined. In the first case, i.e.\ for $\mu^2=-\frac29(z+2)^2$, we indeed see that
\begin{align}
3\lambda_\phi^-=-(z+2).
\end{align}
This continuous ambiguity comes from the appearance of a marginal deformation. Namely, the operator $\phi^3$ is marginal for this specific value of $\mu^2$. In the second case, i.e.\ for $\mu^2= \frac1{16}(z+2-\beta_z)^2-\frac14(z+2)^2$, we find that
\begin{align}
\lambda_\alpha^-+2\lambda_\phi^-\ &=\ -(z+2), & &(1<z<2), \\
\lambda_\alpha^-+2\lambda_\phi^+\ &=\ -(z+2), & &\quad(z>2).
\end{align}
In this case, the operator $(\alpha-\alpha_0)\phi^2$ is marginal.

\subsubsection*{The on-shell action}
For future reference, let us conclude this section with the explicit form of the renormalized on-shell action at the constant level,
\begin{align}
\Gamma\ &=\ \int d^{d+1}x\sqrt{-g}\left(R-2\Lambda-\frac14F_{\mu\nu}F^{\mu\nu}-\frac12m^2A_\mu A^\mu\right)+\int d^{d}\xi\sqrt{-\gamma}\ 2K\nonumber\\
&\quad-\int d^{d+1}x\sqrt{-g}\left(2(z+1)-\frac{z}{2}(\alpha-\alpha_0)+ \frac{z^2(2z-5+\beta_z)}{16(z^2-1)}(\alpha-\alpha_0)^2+\ldots  \right)
\end{align}
The ellipses denote higher-order derivatives as well as terms that are of higher order in $\alpha-\alpha_0$.


\section{Renormalized On-Shell Action}\label{sec:on-shell-action}

Up to this point we did not impose any boundary conditions on our space of solutions. In this section we will analyze the Einstein--Proca equations in the constant perturbation sector up to second order and show that divergences in the third-order on-shell action are indeed removed using our formalism for a large class of boundary conditions.


\subsection{Perturbative analysis of the Einstein--Proca equations}

First, we discuss the first-order solution obtained in \cite{RS} and then we set out to solve the second-order equations. The purpose of finding these solutions is to perform a non-trivial check of the counterterms that we found in \ref{sec:counterterms}. Again, we focus on the non-derivative sector throughout this paper, so we shall restrict our analysis of the linearized field equations to constant modes which only depend on the radial coordinate $r$. For simplicity, we postpone the treatment of vector and tensor perturbations to section \ref{sec:vectortensor}, and here we focus only on the scalar subsector.Furthermore, we set the scalar field to its background value as well, $\phi=0$.

We adopt the same parametrization as \cite{RS} for which the Lifshitz geometry is perturbed as follows.
\begin{align}
\gamma_{tt}\ &=\ -e^{2zr}\Big( 1 + \varepsilon\, f(r) + \varepsilon^2\, \tilde f(r) + \ldots \Big), \\
\gamma_{ij}\ &=\ e^{2r}\delta_{ij}\Big( 1 + \varepsilon\, k(r) + \varepsilon^2\, \tilde k(r) + \ldots \Big), \\
A_{t}\ &=\ \sqrt{-\alpha_0}\,e^{zr}\Big( 1 + \varepsilon\, \big(\,j(r)+\tfrac12f(r)\,\big) + \varepsilon^2\, \big(\,\tilde j(r)+\tfrac12\tilde f(r)\,\big) + \ldots \Big).
\end{align}
We use the small parameter $\varepsilon$ to keep track of the order in the perturbative expansion. We work in radial gauge, which means that the components $g_{r\mu}$ do not receive any corrections.


\textbf{First-order solution.}
As was noted in \cite{RS}, the first order field equations for constant perturbations reduce to the following three equations.
\begin{align}
0&= 2 j'' - (z+1)  f' - (4z+6)  j' + 2(z+4) (z-1) j, \\
0&= (z+1) f''+3(z+1)f'  -(z-1)(4z+2) j' - (z-1)(4z^2+6z+8) j, \\
0&= 2(z+1) k'  +(z+1)  f' + 2(z-1) j' +(z-1)(2z-4) j.
\end{align}
The first two of these equations are second order, while the third one is first order. This means that there must be five integration constants: $c_1, ... , c_5$. The solution is given by
\begin{align}
j &= -\frac{(z+1)c_1}{z-1}\, e^{-(z+2)r} -
\frac{(z+1)c_2}{z-1}\, e^{-\frac{1}{2}(z+2+\beta_z)r} +
\frac{(z+1)c_3}{z-1}\, e^{-\frac{1}{2}(z+2-\beta_z)r}, \\
f &= \frac{4c_1}{z+2}\, e^{-(z+2)r} + 
\frac{2(5z-2-\beta_z)c_2}{z+2+\beta_z}\,
e^{-\frac{1}{2}(z+2+\beta_z)r}   -\frac{2(5z-2+\beta_z)c_3}{z+2-\beta_z}\,
e^{-\frac{1}{2}(z+2-\beta_z)r} + c_4, \nonumber \\
k &=  \frac{2c_1}{z+2}\, e^{-(z+2)r}
-\frac{2(3z-4-\beta_z)c_2}{z+2+\beta_z}\,e^{-\frac{1}{2}(z+2+\beta_z)r}
 +\frac{2(3z-4+\beta_z)c_3}{z+2-\beta_z}\,e^{-\frac{1}{2}(z+2-\beta_z)r} +c_5. \nonumber
\end{align}
This solution holds for the range of the dynamical exponent $1<z<2$ and $z>2$. The case of $z=2$ must be treated separately because of the appearance of logarithmic modes. To see that these logarithmic modes are needed to solve the field equations, notice that e.g.\ the $c_1$ and $c_2$ modes have the same radial behavior when $z=2$ (and similarly for the $c_3$ and $c_{4,5}$ modes). The case of $z=2$ shall be discussed later on, in section \ref{sec:z=2}. It is also interesting to see that the vector modes decouple from the metric modes when $z=1$ (just rescale $c_{1,2,3}\rightarrow (z-1)c_{1,2,3}$ and shift $c_4$ and $c_5$).

For $1< z <2$, all the modes decay as $r\to \infty$; when $z>2$ the $c_{3}$ mode diverges. Notice in particular that $c_{4}$ and $c_{5}$ correspond to linearized diffeomorphisms, generated by the vector field
\begin{equation}
\xi = \frac{c_{4}}{2} t\, \partial_{t} + \frac{c_{5}}{2} x^{i}\, \partial_{i}.
\end{equation}
Since $\alpha$ is a scalar, it cannot depend on $c_{4}$ and $c_{5}$ at linear order. In fact it is easy to see that, at this order
\begin{equation}\label{eq:alpha-j-relation}
\alpha - \alpha_{0} = \varepsilon j(r)+O(\varepsilon^2).
\end{equation}
The $c_1$ mode should be related to the mass of a black hole solution, as suggested by the asymptotically Lifshitz black holes considered for example in \cite{Taylor,TV}. Additional evidence for this is provided by the ADM mass:
\begin{align}
M_\text{ADM} = \varepsilon\frac{4(z-2)}{z+2}\,c_1.
\end{align}
Notice however that this computation is a bit suspicious because the ADM mass is computed by means of background subtraction, and additional counterterms could modify the answer.


\textbf{Second-order solution.}
The second order field equations are given by
\begin{align}
0\ &=\ z(z+2)\,\big(f-2j\big)^2 + \frac{3z-1}{z-1} f'^2 + 4\,j'f' + 4j'^2 + 4z\,jf' + 8z\,jj' + \frac{6z(z+1)}{z-1}\,ff' \nonumber\\
&\quad - 4z\,fj' - \frac{4z(z+2)}{z-1}\,kk'  + \frac{2z}{z-1}\,k'^2 - \frac{2z}{z-1}\,f'k' + \frac{4z}{z-1}\,kk'' + \frac{4z}{z-1}\,ff''\nonumber\\
&\quad +8z(z+2)\,\tilde j - \frac{4z(z+2)}{z-1}\,\tilde f' + 8z\,\tilde j' - \frac{4z(z+2)}{z-1}\,\tilde k' - \frac{4z}{z-1}\,\tilde f'' - \frac{4z}{z-1}\,\tilde k''\\
0\ &=\ z(z+2)\,\big(f-2j\big)^2 + f'^2 + 4\,j'f' + 4j'^2 + 4z\,jf' + 8z\,jj' \nonumber\\
&\quad-2z\,ff' - 4z\,fj' - \frac{24z}{z-1}\,kk'  - \frac{2z}{z-1}\,k'^2 - \frac{8z}{z-1}\,kk'' \nonumber\\
&\quad +8z(z+2)\,\tilde j + 4z\,\tilde f' + 8z\,\tilde j' + \frac{24z}{z-1}\,\tilde k' + \frac{8z}{z-1}\,\tilde k'',\\
0\ &=\ z(z-2)\,\big(f-2j\big)^2 + f'^2 + 4\,j'f' + 4j'^2 + 4z\,jf' + 8z\,jj' \nonumber\\
&\quad-\frac{2z(z+3)}{z-1}\,ff' - 4z\,fj' - \frac{8z(z+1)}{z-1}\,kk' + \frac{2z}{z-1}\,k'^2 +\frac{4z}{z-1}\,f'k' ,\nonumber\\
&\quad +8z(z-2)\,\tilde j + \frac{4z(z+1)}{z-1}\,\tilde f' + 8z\,\tilde j' + \frac{8z(z+1)}{z-1}\,\tilde k'.
\end{align}
Just like the first-order equations, these consist of one first-order differential equation and two second order ones, so again there are five integration constants. Another thing one can read off from these equations is that the only modes that can appear in the second-order functions $\tilde j$, $\tilde f$ and $\tilde k$ are products of the modes we had already found at first order. The solution is thus given by
\begin{align}
\tilde j(r)\ &=\ j_1\,e^{-(z+2)r} + j_2\,e^{-\frac12(z+2+\beta)r} + j_3\,e^{-\frac12(z+2-\beta)r}+j_4 
\nonumber\\ &\quad + j_5\,e^{-(z+2+\beta)r} + j_6\,e^{-(z+2-\beta)r} + j_7\,e^{-2(z+2)r}
\label{eq:2nd-order-sol-j}\\ &\quad + j_8\,e^{-\frac12(3(z+2)+\beta)r} + j_9\,e^{-\frac12(3(z+2)-\beta)r} 		,\nonumber\\
\nonumber\nonumber\\
\tilde f(r)\ &=\ f_1\,e^{-(z+2)r} + f_2\,e^{-\frac12(z+2+\beta)r} + f_3\,e^{-\frac12(z+2-\beta)r}+f_4 \nonumber
\nonumber\\ &\quad + f_5\,e^{-(z+2+\beta)r} + f_6\,e^{-(z+2-\beta)r} + f_7\,e^{-2(z+2)r} 
\label{eq:2nd-order-sol-f}\\ &\quad + f_8\,e^{-\frac12(3(z+2)+\beta)r} + f_9\,e^{-\frac12(3(z+2)-\beta)r},		\nonumber\\
\nonumber\nonumber\\
\tilde k(r)\ &=\ k_1\,e^{-(z+2)r} + k_2\,e^{-\frac12(z+2+\beta)r} + k_3\,e^{-\frac12(z+2-\beta)r} +k_4\nonumber\\ 
&\quad + k_5\,e^{-(z+2+\beta)r}  + k_6\,e^{-(z+2-\beta)r} + k_7\,e^{-2(z+2)r}\label{eq:2nd-order-sol-k}\\
&\quad + k_8\,e^{-\frac12(3(z+2)+\beta)r} + k_9\,e^{-\frac12(3(z+2)-\beta)r}.\nonumber
\end{align}
The coefficients $j_i$, $f_i$, and $k_i$ depend on the dynamical exponent $z$ as well as the first and second-order integration constants $c_j$ and $\tilde c_j$. Instead of listing the coefficients explicitly, we shall discuss which coefficients are fully determined by the field equations and which ones are related by integration constants. The coefficients of the modes that did \emph{not} appear at first order are entirely determined by the field equations and thus only depend on the first-order integration constants $c_i$.\footnote{The coefficients we are talking about here are $j_4$ and $(j_i,f_i,k_i)$ with $i=5,...,9$.} The coefficients $(j_i,f_i,k_i)$ are related by $\tilde c_i$,  with $i=1,2,3$. The coefficients $f_4$ and $k_4$ do not enter the field equations at all, so let us call them $f_4=\tilde c_4$ and $k_4=\tilde c_5$ for consistency of notation.


\subsection{On-shell action}

These solutions allow us to compute the on-shell action as a function of the integration constants up to second order in the expansion parameter $\varepsilon$,
\begin{align}\label{eq:onshell-action_expanded}
\Gamma =S_{(0)}+\varepsilon \,S_{(1)}+\varepsilon^2\, S_{(2)}+\ldots
\end{align}
The leading-order term vanishes, $S_{(0)}=0$, while the first-order term is given by
\begin{equation}
S_{(1)}=\frac{2 c_{1}(z-2) (z+1)}{z+2},
\end{equation}
which reproduces the result of \cite{RS}. At second order in $\varepsilon$ we find
\begin{align}\label{eq:S2}
S_{(2)}&=-\frac{2c_{1} c_{4}\, (z-1)}{z+2}+\frac{2 c_{1} c_{5}\, z (z+1)}{z+2}
+\frac{ 2 c_{2} c_{3}\,(z+1)\big((z-2)\beta -8\big)}{(z-2) (z-1)} + \tilde c_1,
\end{align}
where $\tilde c_{1}$ is a second-order correction to $c_{1}$, which cannot be determined by the asymptotic analysis and is therefore arbitrary. It is pleasing to see that the second order on-shell action is finite. Furthermore we recognize a familiar structure source/state, where $c_{3}$ sources $c_{2}$ while a linear combination of $c_{4}$ and $c_{5}$ sources $c_{1}$. Therefore we identify $c_{3}$, $c_{4}$ and $c_{5}$ with the boundary conditions in the constant scalar perturbations sector. The remaining parameters $c_{1}$ and $c_{2}$ are then naturally identified with the state of the system, and are determined by the initial and final conditions \cite{RS2}.

Since the counterterms in \cite{RS} had been devised to cancel first-order divergences only, one does not expect them to properly renormalize the on-shell action at higher orders. At second order, one finds indeed that the on-shell action is infinite when one uses those counterterms.\\

\textbf{Third-order on-shell action.}
The second order solutions should be sufficient to check finiteness of the on-shell action at third order (we expect only third-order corrections to $c_{1}$ coming from the third order solutions). In fact an explicit computation shows that the third-order contribution $S_{(3)}$ is finite with our counterterms. This provides an additional non-trivial check that our counterterms indeed remove the divergences.

\subsection{Vector and tensor perturbations}
\label{sec:vectortensor}
For vector and tensor perturbations, the linearized analysis should be sufficient to compute the second-order on-shell action. Using again the notation in \cite{RS}, we have:
\begin{align}
\gamma_{ti} &= -e^{2z r} v_{1i}(r) + e^{2 r} v_{2i}(r),\\
\gamma_{ij} &= e^{2 r}  k_{ij}(r),\\
A_{i} &= \sqrt{-\alpha_{0}}\,e^{z r} v_{1i}(r),
\end{align}
where
\begin{equation}
k_{ij}(r) = \left(\begin{array}{cc} t_{d}(r) & t_{o}(r) \\ t_{o}(r) & -t_{d}(r)\end{array}\right).
\end{equation}

\textbf{Vector perturbations.}
The vector sector is parametrized by
\begin{align}
v_{1i}(r) & = c_{1i} + c_{2i}e^{-(z+2) r}+ c_{3i}e^{-3z r},\\
v_{2i}(r) & = \frac{z^{2}-4}{z(z-4)}c_{2i}e^{ (z-4) r}+ \frac{3z}{z+2}c_{3i}e^{-(z+2) r} + c_{4i}.
\end{align}
The on-shell action \eqref{eq:onshell-action_expanded} converges only for $z<4$, and its second-order (in $\varepsilon$) term is given by\footnote{For simplicity, we take the perturbation to lie along the $x-$axis.}
\begin{equation}
S_{(2)} = \frac{4 (z-1)(z+1)}{(z+2)} c_{1i} c_{3i}+\frac{2 (z-1)(z^2-4 z-8)}{z(z-4)}c_{2i} c_{4i}.
\end{equation}
Once again we recognize the source/state structure $c_{1i}/c_{3i}$ and $c_{4i}/c_{2i}$.

\textbf{Tensor perturbations.}
Finally the tensor modes, given by:
\begin{align}
t_{d}(r) & = t_{d1} + t_{d2}e^{-(z+2) r},\\
t_{o}(r) & = t_{o1} + t_{o2}e^{-(z+2) r},
\end{align}
lead to the second-order contribution to the on-shell action \eqref{eq:onshell-action_expanded}
\begin{equation}
S_{(2)} = 2  (z+1) (t_{o1} t_{o2}+t_{d1} t_{d2}).
\end{equation}
The source/state structure is in this case $t_{d1}/t_{d2}$ and $t_{o1}/t_{o2}$.

\subsection{Boundary conditions}

At this point we can draw some conclusions: the modes $c_{4}$, $c_{5}$, $c_{1i}$, $c_{4i}$, $t_{d1}$ and $t_{o1}$ should be interpreted as the sources in the metric sector, because they change the boundary values of $\gamma_{ij}$. The expectation values of the dual operators are given by $c_{1}$, $c_{3i}$, $c_{2i}$, $t_{d2}$ and $t_{o2}$. The mode $c_{3}$ is a source for the massive vector field and is fixed by the leading term in $\alpha-\alpha_{0}$. The corresponding expectation value is given by $c_{2}$.

We note that the mode $c_{2i}$ is problematic when $z>4$, since it leads to a divergent on-shell action.


\subsection{The special case of \textit{z}=2}\label{sec:z=2}
Let us repeat the analysis of the constant (scalar) perturbations for $z=2$. The first-order solution was computed in \cite{RS}; it is given by
\begin{align}
j(r) &= -\left(c_1+c_2\,r\right)\,e^{-4r} +c_3, \nonumber\\
f(r) &= \frac1{12}\left(4c_1-5c_2+4c_2\,r\right)\,e^{-4r} + \left(4c_3\,r+c_4\right), \\
k(r) &= \frac1{24}\left(4c_1+5c_2+4c_2\,r\right)\,e^{-4r} + \left(-2c_3\,r+c_5\right).\nonumber
\end{align}
In this case, the modes $c_1$ and $c_2$ are normalizable, while $c_3$, $c_4$, and $c_5$ are non-normalizable. The possible modes in the second-order solution can again be obtained by squaring the first-order modes. The second-order solution is thus given by
\begin{align}
\tilde j(r) &= \left(j_1+j_2\,r+j_3\,r^2\right)\,e^{-4r} +j_4+j_5\,r+j_6\,r^2 +\left(j_7+j_8\,r+j_9\,r^2\right)\,e^{-8r},\nonumber\\
\tilde f(r) &= \left(f_1+f_2\,r+f_3\,r^2\right)\,e^{-4r} +f_4+f_5\,r+f_6\,r^2 +\left(f_7+f_8\,r+f_9\,r^2\right)\,e^{-8r},\\
\tilde k(r) &= \left(k_1+k_2\,r+k_3\,r^2\right)\,e^{-4r} +k_4+k_5\,r+k_6\,r^2+\left(k_7+k_8\,r+k_9\,r^2\right)\,e^{-8r}.\nonumber
\end{align}
Again, the coefficients $j_i$, $f_i$, $k_i$ depend on the first-order and second-order integration constants $c_j$ and $\tilde c_j$ respectively. We shall not list these coefficients explicitly, but let us mention where the second-order integration constants appear. The coefficients of the modes that were not present in the first-order solution are all fully determined by the field equations.\footnote{To clarify, these coefficients are $j_5$ and $(j_i,f_i,k_i)$ with $i=3,6,7,8,9$.} The coefficients $(j_{1},f_{1},k_{1})$ are related by the integration constant $\tilde c_{1}$, while $(j_{2},f_{2},k_{2})$ are related by $\tilde c_2$, and $(j_4,f_5,k_5)$ are related by $\tilde c_3$. The coefficients $f_4$ and $k_4$ do not enter the field equations at all, so just like the $z\neq2$ case, we call them $f_4=\tilde c_4$ and $k_4=\tilde c_5$ for consistency of notation.

The on-shell action \eqref{eq:onshell-action_expanded} for $z=2$ at first order in $\varepsilon$ is given by
\begin{align}
S_{(1)} = \frac{2c_2}{3},
\end{align}
which reproduces the result in \cite{RS}. At second order, we find
\begin{align}
S_{(2)} = 3 c_{1} c_{3}+\frac{25}{12} c_{2} c_{3}-\frac{1}{6}c_{1} c_{4}-\frac{25}{24} c_{2} c_{4}+c_{1} c_{5}+\frac{19 }{12}c_{2} c_{5}+\tilde c_2,
\end{align}
where $\tilde c_2$ is a correction to the vev $c_2$. We also checked whether the on-shell action is finite at third order and we find that it is.

\textbf{Boundary conditions}
The boundary conditions are essentially the same as in the $z\neq2$ case. The modes $c_3$, $c_4$, $c_5$ should be seen as the sources, while the vevs are represented by $c_1$ and $c_2$. Again, $c_3$ can be interpreted as the source for the vector's mass term via \eqref{eq:alpha-j-relation}. Notice that the $c_2$ mode takes on a similar role as $c_1$ for $z\neq2$, for instance the ADM mass is $M_\text{ADM}=4c_2/3$ when $z=2$.


\section{Conclusions}
We have found a new and systematic method for simultaneously 
determining the boundary conditions on the one hand, and
finding the counterterm action for asymptotically Lifshitz spacetimes on the other hand. 
This method allowed us to find contributions to the `Lifshitz scaling anomaly'. We performed a non-trivial 
consistency check for the counterterms obtained via this method. We saw that our counterterms properly 
renormalize the on-shell action even for higher-order perturbations. The counterterm action we find is 
a local functional of the fields by construction. Moreover, we find that the counterterms are independent of the radial cut-off, unlike some previous approaches, e.g.\ \cite{CHK} for $z=2$. Although
in that paper the counterterms are local functionals of the boundary fields, the coefficients
that appear depend explicitly on the radial cut-off, whereas in our case this cut-off dependence
is only implicit through the dependence of the fields on the radius.
The counterterms found in \cite{RS} only managed to make the on-shell action finite up to linear order. 
In our case, the on-shell action remains finite when we turn on non-normalizable modes.

It should perhaps be emphasized that the idea that one should be able to remove all divergences
using only local counterterms is a conjecture and we have only shown that it works in particular
examples. A full and general proof that this works for all reasonable bulk Lagrangians and to all orders
is lacking.

Although we focused on constant perturbations, the Hamilton--Jacobi analysis can be used to find higher-derivative counterterms as well. See Appendix \ref{sec:higher-derivatives} for an example of such a calculation.
Though our method determines the asymptotic behavior of the fields, there are still some puzzles
that remain. For example, in the computations in section \ref{sec:on-shell-action}, five free parameters appeared, whereas
for a non-degenerate set of equations of motion one would expect to find an even number that can
be split in ``coordinates plus momenta'' or equivalently in sources and expectation values, as was emphasized in \cite{Papadimitriou:2010as}. One
would expect that a canonical analysis in this sector would reveal that one of the five parameters
can be removed by a suitable gauge transformation which seems related to a bulk diffeomorphism (which turns out to be a Lifshitz rescaling at the linear level).
However, a preliminary analysis suggests that the corresponding constraints will end up being non-linear
once higher order corrections are included, and the precise nature of these non-linear boundary
conditions remains to be determined. It is also unclear whether this constraint would somehow
follow from the analysis of the Hamilton-Jacobi equations or require separate input.

There are various qualitative differences for different values of the dynamical exponent $z$. As mentioned in section \ref{sec:on-shell-action}, 
for $z>4$, divergences seem to appear which cannot be canceled using local counterterms.
If we blindly follow the strategy we have been employing, we would be forced to impose more
stringent boundary conditions for $z>4$ which remove these divergences and it would be
interesting to explore this in more detail.
Furthermore, the source for the massive vector field is irrelevant when $z>2$, and therefore we expect that non-local counterterms are needed at sufficiently high order in the sources to make the on-shell action finite, as pointed out in \cite{vanRees:2011fr}.

Another thing which would be interesting to compute is the holographic Weyl anomaly in
a curved background for the Lifshitz case, which from a preliminary analysis seems to involve terms like $R_{ab}A^{a}A^{b}$ for $z=2$ and $d=3$.
Though in principle straightforward, the relevant computations turn out to be extremely 
tedious and we leave this for future work.

One may wonder whether the fact that all divergences can be canceled by local counterterms
is a special feature of field theories with a Lifshitz dual or hold for a more general
class of non-relativistic scale-invariant theories, and it would be interesting to
explore this question directly in field theory.

There are many further directions to explore, such as applications to black hole
solutions, applications to correlation functions, the extension of our work to
Schr\"odinger spacetimes, etc, and we hope to turn back to some of these in due course.


\section*{Acknowledgements}

We would like to thank Geoffrey Comp\`ere, Balt van Rees,   for useful discussions and Dimitrios Korres for 
collaboration at a very early stage of this project.
This work is part of the research programme of the Foundation for Fundamental Research 
on Matter (FOM), which is part of the Netherlands Organisation for Scientific Research (NWO).

\appendix

\newpage
\section{The Einstein--Proca Hamiltonian}\label{sec:hamiltonian}
In this section we shall compute the Hamiltonian associated to the Einstein--Proca action, which is given in section \ref{sec:lsatepa}. We denote by $\Sigma_r$ the surface of constant radial coordinate $r$. We assume for simplicity that $\partial\Sigma_r=\varnothing$, so that we need not worry about possible boundary terms later on. The foliation can be written in the form of a parametric relation $X^\mu=X^\mu(r,x^a)$. It is useful to define the projector
\begin{equation}
p^\mu_a=\frac{\partial X^\mu}{\partial x^a},
\end{equation}
which projects onto the directions tangent to the hypersurface. Thus, the projector is orthogonal to the unit normal, $n_\mu\,p^\mu_a=0$. The cotangent basis is spanned by
\begin{equation}
dX^\mu=r^\mu\,dr+p^\mu_a\,dx^a.
\end{equation}
The vector $r^\mu$ points along the radial flow, which does not necessarily mean it should be proportional to the unit normal. The flow vector is generically given by
\begin{equation}
r^\mu=Nn^\mu+N^a\, p^\mu_a,
\end{equation}
where the normal and tangent pieces are given by the lapse $N$ and shift $N^a$ respectively. The metric $g_{\mu\nu}$ can be rewritten in terms of the fields $(N,N^a,\gamma_{ab})$ as follows.
\begin{equation}
ds^2=N^2\,dr^2+\gamma_{ab}(N^a\,dr+dx^a)(N^b\,dr+dx^b),
\end{equation}
where $\gamma_{ab}=p^\mu_a p^\nu_b\,g_{\mu\nu}$ is the induced metric or first fundamental form. In order to rewrite the gravitational Lagrangian in terms of quantities that are either intrinsic or extrinsic to $\Sigma_r$, we use a projected form of the Gauss--Codazzi equations, namely
\begin{equation}\label{eq:gauss-codazzi}
R^{(d+1)}\ =\ R + K^2 - K_{ab}K^{ab}+2\nabla_\mu\left(n^\nu\nabla_\nu n^\mu-n^\mu\nabla_\nu n^\nu\right).
\end{equation}
The extrinsic curvature, or second fundamental form, is given by $K_{ab}=\frac12\mathscr{L}_n \gamma_{ab}=p^\mu_a p^\nu_b\,\nabla_\mu n_\nu$.\footnote{By the Lie derivative of a tangential object with respect to some $(d+1)$-dimensional vector $\xi^\mu$, e.g. $\mathscr{L}_\xi T_{a_1\cdots a_n}$, we really mean $p^{\mu_1}_{a_1} \cdots p^{\mu_n}_{a_n}\,\mathscr{L}_\xi T_{\mu_1\cdots\mu_n}$.} Similarly, the Maxwell term and the mass term can be split up into normal and tangential pieces using the completeness relation $g^{\mu\nu}=n^\mu n^\nu+p^\mu_a p^\nu_b\,\gamma_{ab}$. Let us use the short-hand notation $\mathcal{V}=n^\mu A_\mu$ and $\mathcal{K}_a=\mathscr{L}_nA_a-\partial_a\mathcal{V}$, such that
\begin{align}
F_{\mu\nu}F^{\mu\nu}\ &=\ F_{ab}F^{ab}+2\mathcal{K}_a \mathcal{K}^a,\\
A_\mu A^\mu\ &=\ A_a A^a + \mathcal{V}^2.
\end{align}
The actions from before thus become\footnote{We take $\Sigma_r$ to be the boundary of our $(d+1)$-dimensional space such that the total-divergence term in \eqref{eq:gauss-codazzi} precisely cancels against the Gibbons--Hawking term in the gravitational action.}
\begin{align}
S_\text{grav}\ &=\ \frac1{2\kappa^2}\int dr\int_{\Sigma_r} d^dx\sqrt{-\gamma}\,N\left(R-2\Lambda + K^2 - K_{ab}K^{ab}\right),\\
S_A\ &=\ \int dr\int_{\Sigma_r} d^dx\sqrt{-\gamma}\,N\bigg( -\frac14F_{ab}F^{ab}-\frac12m^2\,A_a A^a-\frac12\mathcal{K}_a \mathcal{K}^a - \frac12m^2\mathcal{V}^2 \bigg).
\end{align}
Let us define the Lagrangians such that $S_\text{grav}=\int dr\,L_\text{grav}$ and $S_A=\int dr\,L_A$, via which we may obtain the Hamiltonians by means of a Legendre transformation. Before we do so, however, we must define our generalized velocities first.
\begin{align}
\dot\gamma_{ab}\ &=\ \mathscr{L}_r\gamma_{ab}\ =\ 2N\,K_{ab}+2D_{(a}N_{b)}, \\
\dot A_a\ &=\ \mathscr{L}_r A_a\ =\ N\,\mathcal{K}_a+N^bF_{ba}+\partial_a(N\mathcal{V}+N^bA_b).
\end{align}
Finally, the canonical momenta are\footnote{Note that, strictly speaking, the canonical momenta are $\sqrt{-\gamma}\, \pi^{ab}$ and $\sqrt{-\gamma}\, E^a$. This will be taken into account in the Legendre transformations to be performed in \eqref{eq:legendre-transform-grav} and \eqref{eq:legendre-transform-A}.}
\begin{align}
\pi^{ab}\ &=\ \frac1{\sqrt{-\gamma}}\frac{\delta L_\text{grav}}{\delta\dot\gamma_{ab}}\ =\ -\frac1{2\kappa^2}(K^{ab}-K\,\gamma^{ab}), \\
E^a\ &=\ \frac1{\sqrt{-\gamma}}\frac{\delta L_A}{\delta\dot A_a}\ =\ -\mathcal{K}^a.
\end{align}
Now, we are ready to perform the Legendre transformation
\begin{align}
H_\text{grav} &=\! \int_{\Sigma_r} d^dx\sqrt{-\gamma}\, \pi^{ab}\dot\gamma_{ab} - L_\text{grav} \label{eq:legendre-transform-grav}\\
&=\! \int_{\Sigma_r} d^dx\sqrt{-\gamma}\, \left\{N\left[ -2\kappa^2\left(\pi_{ab}\pi^{ab}-\frac1{d-1}\pi^2\right)-\frac1{2\kappa^2}\big(R-2\Lambda\big)\right] + N^a\big(-2D^b\pi_{ab}\big) \right\}
\end{align}
and similarly for the vector field
\begin{align}
H_A &=\! \int_{\Sigma_r} d^dx\sqrt{-\gamma}\, E^a\dot A_a - L_A \label{eq:legendre-transform-A}\\
&=\! \int_{\Sigma_r} d^dx\sqrt{-\gamma}\,\bigg\{ N\left[-\frac12E_a E^a+\frac14F_{ab}F^{ab}+\frac12m^2\,A_aA^a+\frac{m^2}2\,\mathcal{V}^2-\mathcal{V}D_aE^a\right] \nonumber\\
&\hspace{10cm} +N^a\big(F_{ab}E^b-A_aD_bE^b\big) \bigg\}
\end{align}
We can also combine the above two Hamiltonians as
\begin{equation}
H\ =\ H_\text{grav}+H_A\ =\ \int_{\Sigma_r}d^dx\sqrt{-\gamma}\,\Big( N\,\mathcal{H}+N^a\,\mathcal{H}_a \Big),
\end{equation}
where we introduced the Hamiltonian constraint $\mathcal{H}$ and the momentum constraint $\mathcal{H}_a$. After integrating out the non-dynamical field $\mathcal{V}$, the Hamiltonian constraint function is
\begin{equation}
\mathcal{H}\ =\ -2\kappa^2\left(\pi_{ab}\pi^{ab}-\frac1{d-1}\pi^2\right)-\frac12E_a E^a-\frac1{2m^2}(D_aE^a)^2-\mathcal{L}.
\end{equation}
Here, $\mathcal{L}$ is the Lagrangian density restricted to the hypersurface $\Sigma_r$,
\begin{equation}
\mathcal{L}\ =\  \frac1{2\kappa^2}(R-2\Lambda)-\frac14F_{ab}F^{ab}-\frac12m^2\,A_aA^a. 
\end{equation}
The momentum constraint function is given by
\begin{equation}\label{eq:momentum-constraint}
\mathcal{H}_a\ =\ -2D^b\pi_{ab}+F_{ab}E^b-A_a D_bE^b.
\end{equation}


\section{Higher-derivative Counterterms}\label{sec:higher-derivatives}

In this section, we briefly mention how one could systematically solve the local part of the Hamiltonian constraint at the level of higher derivatives (thus finding the local higher-derivative counterterms). We shall put the scalar field $\phi=0$, as it will not be more illuminating in this specific discussion.

\textbf{Local on-shell action Ansatz.}
In the following we are interested in deformations that involve only the metric $\gamma_{ab}$ and the massive vector $A_{a}$.
\begin{equation}
\begin{split}
\mathcal{L}_{\mathrm{loc}} &= U(\alpha)  +\mathcal{C}(\alpha)D_aA^a +\mathcal{D}(\alpha)A^{a}A^{b}D_aA_{b} + \Phi(\alpha) R +\ldots
\end{split}
\end{equation}
Of course, there are other two-derivative terms as well as higher-derivative terms in the Ansatz, but for our purpose of illustrating our method these terms will suffice. We assume that $\partial\Sigma_r=\emptyset$, so that we need to specify the possible counterterms only up to total derivatives. We perform a derivative expansion,
\begin{align}
\mathcal{L}^{(0)}_{\mathrm{loc}} &= U(\alpha),\\
\mathcal{L}^{(1)}_{\mathrm{loc}} &= \mathcal{C}(\alpha)D_aA^a + \mathcal{D}(\alpha)A^{a}A^{b}D_aA_{b},\\
\mathcal{L}^{(2)}_{\mathrm{loc}} &= \Phi(\alpha) R +  \ldots
\end{align}
and
\begin{align}
\mathcal{L}^{(0)} &= -2\Lambda - \frac{m^{2}}{2}\alpha,\\
\mathcal{L}^{(2)} &= R - \frac{1}{4} F_{ab}F^{ab}.
\end{align}
The non-derivative level (level zero) has already been covered in Section \ref{sec:hj}, so let us go directly to the level of one spacetime derivative.

\textbf{One derivative.}
At level one we have only two possible structures,
The canonical momenta are given by
\begin{equation}
\pi^{(1)ab} = \frac1{\sqrt{-\gamma}}\frac{S^{(1)}_\text{loc}}{\gamma_{ab}} =  \left(\frac{1}{2}\mathcal{D}-\mathcal{C}'\right) \left(A^{a} A^{b} (D \cdot A) - 2 A^{c} A^{(a}D^{b)} A_{c} + \gamma^{ab} (A^c A^d D_c A_d)\right),
\end{equation}
and
\begin{equation}
E^{(1)a} = \left(2\mathcal{C}' - \mathcal{D}\right)\left(A^{a} (D \cdot A) - A_{b} D^{a} A^{b}\right).
\end{equation}
The Hamilton constraint can be solved if
\begin{equation}
\label{eqn:level1cond}
\mathcal{D} = 2\mathcal{C}'.
\end{equation}
The resulting term in $\mathcal{L}^{(1)}_{\mathrm{loc}}$ is just a total derivative,
\begin{equation}
\mathcal{C} D_{a} A^{a} + 2\mathcal{C}' A^{a}A^{b}D_{a}A_{b} = D_{a}(\mathcal{C} A^{a}),
\end{equation}
and can be discarded.

\textbf{Two derivatives: the $\Phi R$ term.}
Since the $\Phi R$ term does not mix with the other two-derivative terms, we can consistently solve for $\phi(\alpha)$.
\begin{equation}
\pi^{(2)ab} = \frac{1}{2} \gamma^{ab} \mathcal{L}^{(2)}_{\mathrm{loc}} - \frac{\delta(\Phi R)}{\delta \gamma_{ab}} + \ldots
\end{equation}
\begin{equation}
E^{(2)a} =  2 \Phi' R A^{a} + \ldots
\end{equation}
We now want to compute the coefficient of the $R$ term. Only terms with $R$ or a not contracted $R_{\mu\nu}$ can produce a $R$ term in the final expression:
\begin{align}
\pi_{ab}^{(2)} & = \frac{1}{2} g_{ab} \Phi R - \Phi' A_{a}A_{b} R - R_{ab}\Phi + \ldots,\\
E^{(2)a} & =  2 \Phi' R A^{a} + \ldots
\end{align}
Therefore we have
\begin{equation}
\label{level2R}
2\{S^{(0)}_{\mathrm{loc}},S^{(2)}_{\mathrm{loc}}\}-\mathcal{L}^{(2)} = R\left(-\frac{1}{4}\Phi U + \frac{1}{2}A^{2} (\Phi' U - \Phi U') + (4 A^{2} + A^{4}) \Phi' U' + 1\right) + \ldots = 0.
\end{equation}
Again, we expand $\Phi$ in power series in $(\alpha-\alpha_0)$ where $\alpha=A^2$,
\begin{equation}
\Phi = b_{0} + b_{1} (\alpha -\alpha_{0}) + b_{2} (\alpha  -\alpha_{0})^{2} + \ldots,
\end{equation}
and we plug this result into \eqref{level2R}. We obtain
\begin{equation}
b_{0} = \frac{1}{z}.
\end{equation}
A similar computation for $b_{1}$ yields
\begin{equation}
b_{1} = \frac{5 z- 2 +\beta_{z} }{4 (z+1) (z-2+\beta_{z} )}.
\end{equation}
There does not seem to be a continuous ambiguity for the higher order coefficients. 

Let us briefly discuss an important feature of \eqref{level2R}. The function $\Phi$ satisfies a first-order differential equation, therefore it seems somewhat strange that we were able to determine the coefficient $b_{0}$ uniquely, which amounts to specifying the initial condition. The reason for this is that, since we want to compute the polynomial part of the on-shell action, we are using a power-series expansion. Nevertheless, the general solution of the differential equation might not be polynomial, so by requiring that our solution is a polynomial, we are effectively determining the initial condition. We illustrate this phenomenon with a toy example. Consider the differential equation:
\begin{equation}
x f'(x) + a f(x) + 1 = 0.
\end{equation}
If $a\neq 0$ this has the following general solution:
\begin{equation}
-\frac{1}{a} + A x^{-a},
\end{equation}
where $A$ is an arbitrary constant. If $a\neq 0, -1,-2,\ldots$, then the solution is not polynomial and using a Taylor expansion amounts to choosing $A=0$. Nevertheless, if the coefficient $a$ is a negative integer, the solution is indeed a polynomial but $A$ is undetermined. This amounts precisely to a continuous ambiguity that we would find by using the power-series method.

Equation \eqref{level2R} can be cast in a form similar to the toy model we just considered:
\begin{equation}
\left((\alpha^{2} + 4\alpha)  U'  + \frac{1}{2} \alpha U \right) \Phi' + \left(\frac{1}{2}\alpha U' - \frac{1}{4} U\right)\Phi + 1 = 0.
\end{equation}
The coefficient of $\Phi'$ is simply $-\partial_{r} \alpha$, and stability required that this coefficient vanishes as $\alpha \to \alpha_{0}$, as we explained at the end of section \ref{acbycs}. This feature is very general and it explains why the HJ method is able to fix the derivative counterterms.

\newpage
\bibliographystyle{JHEP}
\bibliography{references}
\end{document}